\documentclass[dvips]{article}
\usepackage{lscape,epsfig}
\usepackage{amssymb}
\usepackage{amsmath}
\usepackage{rotating}
\usepackage{graphicx}

\textheight=22cm \DeclareSymbolFont{ppa}{OT1}{ppl}{m}{it}
\DeclareSymbolFont{ppa}{OT1}{ppl}{m}{it}
\DeclareMathSymbol{\vv}{\mathalpha}{ppa}{'166}

\begin{document}

\newcommand{\dd}{\,{\rm d}}
\newcommand{\ie}{{\it i.e.},\,}
\newcommand{\etal}{{\it et al.\ }}
\newcommand{\eg}{{\it e.g.},\,}
\newcommand{\cf}{{\it cf.\ }}
\newcommand{\vs}{{\it vs.\ }}
\newcommand{\zdot}{\makebox[0pt][l]{.}}
\newcommand{\up}[1]{\ifmmode^{\rm #1}\else$^{\rm #1}$\fi}
\newcommand{\dn}[1]{\ifmmode_{\rm #1}\else$_{\rm #1}$\fi}
\newcommand{\upd}{\up{d}}
\newcommand{\uph}{\up{h}}
\newcommand{\upm}{\up{m}}
\newcommand{\ups}{\up{s}}
\newcommand{\arcd}{\ifmmode^{\circ}\else$^{\circ}$\fi}
\newcommand{\arcm}{\ifmmode{'}\else$'$\fi}
\newcommand{\arcs}{\ifmmode{''}\else$''$\fi}
\newcommand{\MS}{{\rm M}\ifmmode_{\odot}\else$_{\odot}$\fi}
\newcommand{\RS}{{\rm R}\ifmmode_{\odot}\else$_{\odot}$\fi}
\newcommand{\LS}{{\rm L}\ifmmode_{\odot}\else$_{\odot}$\fi}

\newcommand{\Abstract}[2]{{\footnotesize\begin{center}ABSTRACT\end{center}
\vspace{1mm}\par#1\par
\noindent
{~}{\it #2}}}

\newcommand{\TabCap}[2]{\begin{center}\parbox[t]{#1}{\begin{center}
  \small {\spaceskip 2pt plus 1pt minus 1pt T a b l e}
  \refstepcounter{table}\thetable \\[2mm]
  \footnotesize #2 \end{center}}\end{center}}

\newcommand{\TableSep}[2]{\begin{table}[p]\vspace{#1}
\TabCap{#2}\end{table}}

\newcommand{\FigCap}[1]{\footnotesize\par\noindent Fig.\  %
  \refstepcounter{figure}\thefigure. #1\par}

\newcommand{\TableFont}{\footnotesize}
\newcommand{\TableFontIt}{\ttit}
\newcommand{\SetTableFont}[1]{\renewcommand{\TableFont}{#1}}

\newcommand{\MakeTable}[4]{\begin{table}[htb]\TabCap{#2}{#3}
  \begin{center} \TableFont \begin{tabular}{#1} #4
  \end{tabular}\end{center}\end{table}}

\newcommand{\MakeTableSep}[4]{\begin{table}[p]\TabCap{#2}{#3}
  \begin{center} \TableFont \begin{tabular}{#1} #4
  \end{tabular}\end{center}\end{table}}

\newenvironment{references}%
{
\footnotesize \frenchspacing
\renewcommand{\thesection}{}
\renewcommand{\in}{{\rm in }}
\renewcommand{\AA}{Astron.\ Astrophys.}
\newcommand{\AAS}{Astron.~Astrophys.~Suppl.~Ser.}
\newcommand{\ApJ}{Astrophys.\ J.}
\newcommand{\ApJS}{Astrophys.\ J.~Suppl.~Ser.}
\newcommand{\ApJL}{Astrophys.\ J.~Letters}
\newcommand{\AJ}{Astron.\ J.}
\newcommand{\IBVS}{IBVS}
\newcommand{\PASP}{P.A.S.P.}
\newcommand{\Acta}{Acta Astron.}
\newcommand{\MNRAS}{MNRAS}
\renewcommand{\and}{{\rm and }}
\section{{\rm REFERENCES}}
\sloppy \hyphenpenalty10000
\begin{list}{}{\leftmargin1cm\listparindent-1cm
\itemindent\listparindent\parsep0pt\itemsep0pt}}%
{\end{list}\vspace{2mm}}

\def\TYLDA{~}
\newlength{\DW}
\settowidth{\DW}{0}
\newcommand{\dw}{\hspace{\DW}}

\newcommand{\refitem}[5]{\item[]{#1} #2%
\def\REFARG{#3}\ifx\REFARG\TYLDA\else, {\it#3}\fi
\def\REFARG{#4}\ifx\REFARG\TYLDA\else, {\bf#4}\fi
\def\REFARG{#5}\ifx\REFARG\TYLDA\else, {#5}\fi.}

\newcommand{\Section}[1]{\section{#1}}
\newcommand{\Subsection}[1]{\subsection{#1}}
\newcommand{\Acknow}[1]{\par\vspace{5mm}{\bf Acknowledgements.} #1}
\pagestyle{myheadings}

\newfont{\bb}{ptmbi8t at 12pt}
\newcommand{\xrule}{\rule{0pt}{2.5ex}}
\newcommand{\xxrule}{\rule[-1.8ex]{0pt}{4.5ex}}
\def\thefootnote{\fnsymbol{footnote}}
\begin{center}

{\Large\bf
An unusual eclipsing blue straggler V8-NGC~6752
}

\vskip1cm { \large J.~~K~a~l~u~z~n~y$^{1}$,~~M.~~R~o~z~y~c~z~k~a$^{1}$,
           ~~I.~B.~~T~h~o~m~p~s~o~n$^{2}$~~and~~K.~~Z~l~o~c~z~e~w~s~k~i$^{1}$ 
       \vskip3mm {$1$ Nicolaus Copernicus Astronomical Center,\\
        ul. Bartycka 18, 00-716 Warsaw, Poland\\
        e-mail: (jka;mnr;kzlocz)@camk.edu.pl\\
      $^{2}$Carnegie Observatories, 813 Santa Barbara Street,
      Pasadena, CA 91101, USA \\
      e-mail: ian@obs.carnegiescience.edu\\
}
}
\end{center}

\Abstract{We report the analysis of a binary blue
straggler in NGC 6752 with a short orbital period of 0.315 d and a
W UMA-type light curve. We use photometric data spanning 13 years 
to place limits on the mass ratio ($0.15\lesssim q 
\lesssim 0.35$), luminosity ratio ($L_{1}/L_{2}\approx 4.0$) and the
ratio of the radii of the components ($r_{1}/r_{2}\approx 2.0$). The
effective temperatures of the components are nearly identical, and
the system is detached or semi-detached (in the latter case the
component filling its Roche lobe is the secondary). Such a
configuration is unusual given the shortness of the orbital 
period, and it must have resulted from  substantial mass exchange.
We suggest that some secondaries of W UMa-type stars, 
normally regarded as main sequence 
objects which fill their Roche lobes to different degrees, in fact 
may be shell-burning cores of originally more massive components.
}

{\bf Key words:} {\it
blue stragglers --
binaries: eclipsing -- globular clusters: individual (NGC~6752)}

%___________________________________________________________________________

\section{Introduction} \label{sect:intro}

Blue stragglers (BSs) occupy the extension of the main sequence
above the turnoff point in color-magnitude diagrams of parent
stellar populations. Since their detection in M3 (Sandage 1953) BSs
have been identified in dozens of globular and open clusters. Two
main mechanisms responsible for the formation of these objects have
been advocated -- mass transfer in close binaries (McCrea 1964), and
mergers resulting from direct collisions between stars in dense
environments (Benz \& Hills 1987). Knigge et al. (2009) found that
the number of BSs belonging to a given globular cluster (GC) scales
nearly linearly with the core mass of that cluster but is nearly 
independent of the core radius. This implies that the main channel 
leading to the formation of BSs is binarity. Moreover, Perets 
\& Fabrycky (2009) argue that the progenitors of BSs form in primordial 
triple stars through the Kozai mechanism (Kozai 1962). 
This is in line with the 
observational evidence that most (possibly all) close binaries are 
formed in triple systems (Tokovinin et al. 2006).

The sample of candidate BSs in GCs includes several variables with
light curves typical of contact binaries. The discovery of two W~UMa
type systems in NGC~5466 (Mateo et al. 1990) was followed by the
detection of further such systems in other clusters, mostly by the OGLE
and CASE groups (see Rucinski 2000). Surprisingly,
until now not a single mass determination or even a reliable light
curve solution has been obtained for a GC contact binary. In fact,
masses resulting from the analysis of light and radial velocity
curves have been derived for just two binary BSs from GCs (Kaluzny
et al. 2007a, 2007b). Some attempts have been made to estimate masses of
apparently single BSs in GCs based on the modeling of spectra (\eg 
De Marco et al. 2005) or pulsation periods of SX~Phe variables 
(Gilliland 1998). However, the actual accuracy of these
determinations is hard to estimate.

The eclipsing binary V8-NGC~6752 (hereafter V8) was discovered by
Thompson et al. (1999) during a photometric survey of the cluster
field. The system has a W~UMa-type light curve, and an orbital
period of 0.31~d. Our new photometry (Kaluzny \& Thompson 2009)
revealed that the secondary eclipse of V8 is total. As it was first
discussed by Mochnacki \& Doughty (1972), light curves of contact
binaries showing total eclipses can be uniquely solved, yielding
the mass ratio among other parameters. This is in contrast to
contact systems with partial eclipses whose light curves usually
cannot be reliably solved without spectroscopic information about
the mass ratio.\footnote{Except for relatively rare systems with 
mass ratios close to unity.} In short, the totality of the primary
eclipse is sufficient to determine reliable apparent magnitudes of
the individual components.

If V8 is indeed a contact binary belonging to the cluster then one
can estimate its absolute parameters based solely on the
photometric information, which together with the known distance to the cluster
would yield the absolute magnitudes of its components. Then,
based on effective temperatures estimated from colors, one can
derive absolute radii of both stars. Knowing absolute radii and
relative radii (the latter from the light curve solution) 
the absolute size of the orbit can then be derived,  yielding the total
mass of the system from Kepler's law. Given the paucity of reliable
determinations of masses for BSs in GCs we decided to analyze the
light curves of V8 hoping to estimate its parameters. To
our surprise it turned out that the binary is most likely a detached
system with a mass ratio $q<0.35$. This is an unexpected
result given the short orbital period of the system and the similar
effective temperatures of its components.

\section{Observational properties of V8}

\subsection{The data}
Photometric data used in this study are based on CCD images obtained
at Las Campanas Observatory with the 2.5-m du Pont and the 1.0-m
Swope telescopes. The du Pont data collected in 1998 season were
described by Kaluzny \& Thompson (2009). They have been supplemented
with observations from 2007, 2008 and 2009 seasons. The Swope data
collected in 1996 and 1997 seasons were described by Thompson et al.
(1999). They have been supplemented with observations from the 2007
observing season.

Differential photometry was extracted using image subtraction
techniques described by Kaluzny \& Thompson (2009). We also
re-reduced 1996 and 1997 Swope images analyzed earlier by Thompson
et al. (1999). Differential light curves extracted with this method
are free from systematic errors due to crowding. However, this holds
only for light curves expressed in differential counts. Conversion
of a given light curve from differential counts to stellar
magnitudes requires determination of the pedestal flux for a target
star on the reference image. At this stage a failure to correct for 
close visual companion(s) to the target leads to systematic errors
in the light curve expressed in magnitudes. V8 is
present on some HST-WFPC2 images available from the MAST archive
(Program 8256; PI W. Landsman). An examination of these images
revealed two close visual companions to V8 at angular distances of
about 1 arcsec but we found no evidence that the profile of V8
itself was affected by an unresolved blend. The two close companions
posed no problem as they were resolved also in our ground based
data.

%In Fig. 1 we show the
%area around V8 extracted from one of these images. All close
%visual companions visible near the variable were resolved
%on the reference $BV$ images used for processing of duPont frames.
%As for data obtained with the Swope telescope we have
%used a complete list

\subsection{Period Study}\label{sect:Period}
We measured 17 times of eclipses for V8 from the available data;
their values, along with errors determined using the method of Kwee
\& Van Woerden (1956) are given in Table 1. The $O-C$ values listed
in the table correspond to the linear ephemeris
\begin{equation}
 {\rm Min(I)} = {\rm HJD}~2~450~961.75164(9) + 0.31491223(2)
\end{equation}
determined from a least squares fit to the data. Formal errors of
times of minima returned by Kwee \& Van Woerden algorithm were
increased by a factor of 3.0 to assure that the reduced $\chi^{2}$ of the
fit is equal to unity. The linear ephemeris provides a good fit and
there is no evidence for any detectable period change during the
interval 1996--2009 covered by the observations. As there is a large
gap in the observations between 1997 and 2007, this result should be
viewed with some caution. However, observations analyzed separately
for the 1996-1998 and 2007-2009 time intervals give the same orbital
period. Most W UMa-type stars show noticeable changes of orbital
period with $dP/dt \approx 10^{-5}$, resulting in O--C deviations of
the order of 0.01~d on a time base of a few years (Kreiner et al.
2000; Rucinski 1985). Also, noticeable changes of orbital periods are
usually found in short-period algols. In this context the constancy
of the orbital period of V8 is exceptional.

\begin{table}
\begin{center}
\caption{Times of Minima and $O-C$ Values for V8
}
\begin{tabular}{rccc}
\hline
  Cycle & $T_0$& Error  & $O-C$ \\
\hline
      -2217.0  &     263.59120  &       0.00060  &       0.00004\\
      -2216.5  &     263.74830  &       0.00300  &       0.00039\\
      -1130.0  &     605.90160  &       0.00090  &      -0.00077\\
      -1127.5  &     606.68900  &       0.00060  &      -0.00089\\
      -1127.0  &     606.84650  &       0.00090  &      -0.00093\\
       -994.0  &     648.72850  &       0.00090  &       0.00039\\
       -810.5  &     706.51550  &       0.00120  &      -0.00021\\
          0.5  &     961.90910  &       0.00024  &       0.00000\\
          3.5  &     962.85336  &       0.00018  &       0.00048\\
         13.0  &     965.84528  &       0.00049  &       0.00023\\
         19.5  &     967.89278  &       0.00019  &      -0.00034\\
         22.5  &     968.83727  &       0.00022  &      -0.00010\\
      10568.0  &    4289.74410  &       0.00270  &      -0.00001\\
      10568.5  &    4289.90230  &       0.00090  &      -0.00076\\
      12730.5  &    4970.74223  &       0.00033  &      -0.00045\\
      12839.0  &    5004.90964  &       0.00033  &       0.00012\\
      12864.0  &    5012.78232  &       0.00027  &       0.00025\\
\hline
\end{tabular}
\end{center}
\end{table}

\subsection{Light curves}
In Fig. 1 we show $V$ and $B-V$  curves of the du Pont photometry of
V8 phased according to the ephemeris given in equation (1).  
We note that these observations span 11 years, and yet
they were obtained with the same instrumental setup with the only
inhomogeneity due to changes in reflectivity of the telescope
optics. The combined light curve is stable with
little evidence for any seasonal changes of its shape. The depths of
both minima as well as the level of maximum light were the same
during all four observing seasons (1998, 2007, 2008 and 2009). Such
stability of the light curve is very unusual for W UMa stars
(Rucinski 1985). In fact, seasonal instability  of light curves is
common for all short period close binaries whose components
have convective envelopes. It is believed that magnetic
activity, and stellar spots in particular, are responsible for
the observed changes. The data suggest that V8 shows little magnetic activity
and that its components have radiative envelopes. This conclusion is
further supported by the exceptional symmetry of the light curve.
Apparently during periods covered by our observations no large spots
were present on the photospheres of the V8 components.

The totality phase of the secondary eclipse of this system lasts for
about $0.07~P$. Figure 2 presents in detail three well sampled
secondary eclipses from the 1998 season. The primary eclipse is only
marginally deeper than the secondary one, and the color of V8
changes almost inappreciably with orbital phase. Apparently the
effective temperatures of the components are very similar .

\subsection{Membership Status}
NGC~6752 is located at a relatively high galactic latitude of $b=
-25.^{\circ}6$. Therefore the population of the central region of
its field must be strongly dominated by member stars. V8 is seen 
at a projected distance $d=2.0$~arcmin from the cluster center. For 
comparison, the half mass radius of NGC~6752 is $r_{h}=2.34$~arcmin 
(Harris 1996). Thus there is a good chance that the system  
belongs to the cluster. However, a definitive assessment of the 
membership status has to wait until its proper motion or systemic 
radial velocity is determined. 

We are planning to use our data to perform a detailed proper motion 
study of the cluster. For the moment we can only provide qualitative 
evidence in favor of the membership of V8. As  discussed by Eyer 
\& Wo\'zniak (2001), the image subtraction technique can be used to 
detect high proper motion stars located in very dense stellar fields.
Most stars at the center of the cluster 
field should show uniform displacements caused by the proper motion 
of NGC~6752 as a whole. In contrast, unrelated field objects (given 
the cluster's location -- mostly foreground stars) should exhibit 
noticeable displacements with respect to the background population.
%As a result characteristic dipoloidal residuals can be  seen
%on differential images at positions of field stars.
This is illustrated in Fig. 3 which shows the reference $V$ frame
for the 1998 data together with a differential image obtained by subtracting the
reference frame from an average of several individual exposures taken
during the 2009 season. At the location of V8 only a symmetric residual 
is seen, resulting entirely from the variability of the star. Most
stars leave no significant trace on the differential image.
Characteristic bipolar residuals can be seen, however, for a few
stars (likely field objects) displaced with respect to the cluster. We
conclude that V8 is likely to share the proper motion of NGC~6752, which
would mean that it belongs to the cluster.

\section{Light curve solution}\label{sect:vars}

We analyzed the light curve of V8 using the Wilson-Devinney code
(Wilson \& Devinney 1971) as implemented in PHOEBE V29.d package
(Pr\v{s}a \& Zwitter 2005). The following parameters were adjusted:
orbital inclination $i$, effective temperature of the secondary
$T_{2}$, gravitational potentials $\Omega_{1}$ and $\Omega_{2}$, and
the relative luminosity $L_{1}/L_{2}$. The mass ratio $q$ and the
dimensionless potentials $\Omega_{1,2}$ translate into the relative
radii of the components $r_{1,2}$. The temperature of the primary,
$T_{1}$, was determined from the dereddened color index $(B-V)_{1}$
using the calibration of Worthey \& Lee (2006). We adopted an
interstellar reddening of $E(B-V)=0.04$ and a cluster metallicity $[{\rm
Fe/H}]=-1.56$ (Harris 1996). During the totality phase of the
secondary eclipse we measured $(B-V)_1=0.332$, leading to $T_{1}=6990$.
This value places the primary between cool stars with convective
envelopes and  hotter stars with radiative envelopes. Since no photospheric 
activity is detected in our data, we assumed that the components of V8 have
radiative envelopes (see Sec. 3.3). Accordingly, we adopted
bolometric albedos $A_{1,2}=1.0$ and gravity darkening coefficients
$g_{1,2}=1.0$.

For contact binaries with total eclipses the duration of totality
together with the observed amplitude of light variation provides
tight constraints on $q$ and $i$. Using Figure 3 from Mochnacki \&
Doughty (1972) we find that for a contact configuration the mass
ratio of V8 cannot be larger than 0.30. In principle, for the allowed 
range of $q$, we only needed to find a unique pair ($i$, $q$) which 
would reproduce both the length of
totality and the amplitude of the light curve.
%{\bf For
%$\Omega \approx \Omega _{in}$ the observed
%amplitude of the light curve and duration of totality of secondary
%eclipse can be reproduced for  $q\approx $ and $i\approx $ .}
However, it turned out that no such pair provides a
satisfactory fit to the whole light curve as long as the contact 
configuration is enforced. All trial solutions with fixed
$q$ evolved to a detached configuration. At this point we decided to
obtain a grid of solutions for several values of $q$. Only the $V$
light curve was considered -- it has a better quality
and contains more points than the $B$ curve. Including the $B$ curve
would have little impact on the results, as both components have a
very similar surface brightness.

The  grid of solutions is presented in Table 2 (the 
radii are geometrical averages of four values returned by the W-D
code for the different axes of the components). Satisfactory fits to the
light curve could be obtained for $0.15 \lesssim q \lesssim 0.30$.
Fits for $q<0.15$ and for $q>0.30$ show systematic residuals for
some sections of the light curve (in particular, for $q<0.15$ the
synthetic light curves have amplitudes that are too small). 
Solutions for $q>0.16$ imply a detached configuration. With
decreasing $q$ the secondary component fills larger and larger
fraction of its Roche lobe, and finally at $q=0.15$ a semi-detached
configuration is reached. The radius of the primary component is
smaller than the radius of its Roche lobe by 11\%, 15\% and 19\% for
$q=0.30$, $q=0.20$ and $q=0.15$, respectively. The quality of the
fit with $q=0.20$ is illustrated in Fig. 4.
% which shows residuals between observed and
%synthetic $BV$ light curves.
For this value of $q$ a simultaneous solution for $BV$ curves 
was obtained in order to find the luminosity ratios in both 
filters.

\begin{table}\label{tab:radenv}
\begin{center}
\caption{Light curve solutions for V8 for several values of the mass
ratio $q$ (kept fixed during iterations). Radiative envelopes are
assumed for both components.}
\begin{tabular}{llllllll}
\hline
$q$ & $rms^{a}$ & $i$ & $T_{2}$ & $(L_{1}/L_{2})_{V}$ &$<r_{1}>$ & $<r_{2}>$ & Conf$^{b}$\\
\hline
3.00 & 0.0353 &90.00 &7055 &7.746 &0.278 &0.099 &D\\
2.00 & 0.0309 &90.00 &6961 &6.900 &0.313 &0.119 &D\\
1.70 & 0.0289 &90.00 &6942 &6.552 &0.328 &0.129 &D\\
1.45 & 0.0271 &90.00 &6936 &6.111 &0.343 &0.139 &D\\
1.20 & 0.0234 &90.00 &6921 &6.011 &0.361 &0.148 &D\\
1.05 & 0.0219 &90.00 &6891 &5.884 &0.373 &0.155 &D\\
0.95 & 0.0205 &90.00 &6892 &5.778 &0.380 &0.159 &D\\
0.85 & 0.0190 &90.00 &6883 &5.640 &0.390 &0.166 &D\\
0.75 & 0.0171 &90.00 &6868 &5.475 &0.401 &0.174 &D\\
0.65 & 0.0149 &90.00 &6848 &5.373 &0.413 &0.182 &D\\
0.60 & 0.0138 &90.00 &6843 &5.291 &0.419 &0.187 &D\\
0.55 & 0.0127 &90.00 &6836 &5.190 &0.425 &0.192 &D\\
0.50 & 0.0116 &90.00 &6834 &5.075 &0.432 &0.198 &D\\
0.45 & 0.0104 &90.00 &6835 &4.969 &0.439 &0.203 &D\\
0.40 & 0.0094 &90.00 &6840 &4.844 &0.446 &0.209 &D\\
0.35 & 0.0085 &90.00 &6849 &4.672 &0.453 &0.216 &D\\
0.30 & 0.0080 &89.93 &6875 &4.484 &0.459 &0.223 &D\\
0.275& 0.0780 &89.93 &6892 &4.389 &0.461 &0.226 &D\\
0.25 & 0.0076 &89.86 &6907 &4.264 &0.463 &0.230 &D\\
0.225& 0.0075 &89.61 &6948 &4.146 &0.464 &0.232 &D\\
0.20 & 0.0075 &89.27 &6980 &3.973 &0.465 &0.235 &D\\
0.185& 0.0075 &90.00 &7012 &3.936 &0.464 &0.236 &D\\
0.175& 0.0076 &90.00 &7033 &3.880 &0.464 &0.237 &D\\
0.17 & 0.0077 &90.00 &7046 &3.853 &0.464 &0.237 &D\\
0.16 & 0.0077 &90.00 &7071 &3.795 &0.463 &0.238 &D\\
0.15 & 0.0080 &90.00 &7045 &3.735 &0.464 &0.247 &SD2\\
0.14 & 0.0081 &90.00 &7109 &3.699 &0.461 &0.244 &SD2\\
0.13 & 0.0085 &90.00 &7177 &3.633 &0.457 &0.238 &SD2\\
0.12 & 0.0100 &90.00 &7249 &3.543 &0.450 &0.233 &SD2\\
0.11 & 0.0128 &90.00 &7333 &3.505 &0.449 &0.227 &SD2\\
\hline
\end{tabular}
\end{center}
\vspace*{-0.3cm} {\footnotesize Note: $^{a}$ - $rms$ residual of the
fit; $^{b}$ - D and SD2 indicate detached system and semi-detached 
system with the secondary component filling its Roche lobe. }
\end{table}

\begin{table}
\begin{center}
\caption{Same as in Table 2, but for components with convective envelopes.}
\begin{tabular}{llllllll}
\hline
$q$ & $rms^{a}$ & $i$ & $T_{2}$ & $(L_{1}/L_{2})_{V}$ &$<r_{1}>$ & $<r_{2}>$ & Conf$^{b}$\\
\hline
0.60  &0.0121 &90.00 &7184 &4.388   &0.4470 &0.1943 &D\\
0.55  &0.0108 &90.00 &7180 &4.383   &0.4563 &0.1986 &D\\
0.50  &0.0097 &90.00 &7155 &4.292   &0.4604 &0.2050 &D\\
0.45  &0.0087 &90.00 &7138 &4.221   &0.4653 &0.2109 &D\\
0.40  &0.0080 &90.00 &7126 &4.133   &0.4706 &0.2174 &D\\
0.35  &0.0077 &90.00 &7110 &4.022   &0.4751 &0.2248 &D\\
0.35  &0.0077 &86.74 &7108 &4.011   &0.4770 &0.2259 &D\\
0.30  &0.0077 &88.36 &7092 &3.881   &0.4779 &0.2331 &D\\
0.30  &0.0076 &85.63 &7096 &3.876   &0.4779 &0.2331 &D\\
0.275 &0.0077 &89.11 &7091 &3.802   &0.4810 &0.2460 &D\\
0.275 &0.0076 &85.29 &7090 &3.794   &0.4787 &0.2370 &D\\
0.225 &0.0078 &89.99 &7095 &3.640   &0.4760 &0.2479 &D\\
0.225 &0.0077 &85.84 &7098 &3.638   &0.4775 &0.2442 &D\\
0.185 &0.0077 &86.24 &7102 &3.480   &0.4798 &0.2604 &D\\
0.17  &0.0076 &90.02 &7115 &3.442   &0.4700 &0.2535 &SD2\\
0.17  &0.0077 &86.01 &7131 &3.443   &0.4728 &0.2535 &SD2\\
0.17  &0.0076 &86.85 &7123 &3.443   &0.4715 &0.2535 &SD2\\
0.16  &0.0079 &88.95 &7152 &3.416   &0.4669 &0.2510 &SD2\\
0.15  &0.0087 &88.95 &7200 &3.396   &0.4641 &0.2479 &SD2\\
0.14  &0.0103 &89.00 &7250 &3.367   &0.4593 &0.2430 &SD2\\
\hline
\end{tabular}
\end{center}
\vspace*{-0.3cm} {\footnotesize Note: $^{a}$ - $rms$ residual of the
fit; $^{b}$ - D and SD2 indicate detached system and semi-detached 
system with the secondary component filling its Roche lobe. }
\end{table}

One may wonder if a contact configuration could be fitted to the
observations when convective envelopes are assumed for both
components. Such envelopes imply bolometric albedos $A_{1,2}=0.5$
and gravity darkening coefficients $g_{1,2}=0.32$. A grid of
solutions obtained for such a case is presented in Table 3. The
results are very similar to those obtained earlier for the radiative
case. The only difference it that satisfactory fits with a
semi-detached configuration can now be obtained for a slightly wider
range of $q$.

Finally we attempted to fit the light curve allowing for the
presence of a third light ($l_{3}$). As mentioned above, the HST
image does not indicate any blending, nor there is any
evidence for a light-time effect in the timing of the eclipses of V8.
However, we cannot a priori exclude with
certainty the possibility that the image of V8 is contaminated by an
unresolved third star. After numerous trials we found that allowing
for a third light in the system does not influence the results: 
independently of the assumed starting value of $q$ the solutions 
quickly evolved toward $l_{3}=0$. The same applies also to trial 
solutions with contact configurations.

We conclude that the V8 system is either detached or semi-detached
(in the later case the component filling its Roche lobe is the secondary), 
and its mass ratio is low ($0.15<q<0.35$). Tables 2 \& 3 indicate 
that the ratio of radii is $r_{1}/r_{2}\approx 2$, and the  ratio of
$V$-band luminosities is $L_{1}/L_{2}\approx 4$ (or 3.4 in the
"convective" case).

%
%This is quite understandable
%given shape of the light curve. It shows a long lasting total eclipse what
%implies high inclination and noticeable difference of dimensions of components.
%At the same time strong curvature of light curve outside of eclipses
%requires that at lest primary component is close to filling its Roche lobe.
%Addition of third light results in flattening of synthetic light curves.

%Can we use information about distance to select most likely
%solution by deriving absolute parameters?
%Use VMinII and radius-back.

\section{Summary and Discussion}\label{sect:summary}

Our analysis indicates that, despite its short orbital period and
W~UMa-type light curve, the binary V8 is a detached or semidetached
system. This is an unexpected finding, as non-contact systems
are extremely rare among non-degenerate binaries with periods $P\leq
0.45$~d (Hilditch et al. 1988). To the best of our knowledge the
only known systems of this kind are V361~Lyr (Hilditch et al.
1997; Kaluzny 1991; $P=0.30~d$) and OGLE-BW3-V38 (Maceroni \&
Rucinski 1997; $P=0.20$~d). The first of these is a semi-detached system
with the more massive component filling its Roche lobe, while the
latter is probably a detached system composed of two M dwarfs. Another 
possible non-contact binary of this kind is W~Crv (Rucinski
\& Lu 2000), however its actual configuration is poorly
constrained. We note that V8 cannot be considered as an
example of a system caught in the brief contact-break predicted
by ``the thermal relaxation oscillations theory" (Lucy \& Wilson 1979).
In such a system the {\em primary} of V8 would have to fill its
Roche lobe, and the components should have very different effective
temperatures (the secondary being cooler). This is certainly
excluded.

The stability and symmetry of the light curve of V8 together with the
apparent constancy of the orbital period favors a detached
configuration. Despite the totality of one of the eclipses the
photometric analysis does not allow an accurate determination of the mass
ratio of the system. However, it strongly favors the range
$0.15\lesssim q \lesssim 0.35$. The components differ significantly
in size ($r_{1}/r_{2}\approx 2$), yet they have very
similar effective temperatures. Figure 5 shows the location of the
whole system and each of its components separately on the
color-magnitude diagram of NGC~6752. When calculating the
individual magnitudes of the components we adopted a luminosity
ratio of 3.97, the value implied by the solution with $q=0.20$ for
radiative envelopes. Note, however, that the
derived value of $L_{1}/L_{2}$ depends rather weakly on the assumed
$q$ (see Tables 2 and 3).

It is worth noting that the primary of V8 is located right at the
extension of the unevolved main sequence of the cluster. This gives
one more argument for the cluster membership of the binary, and
suggests that the primary has the properties of an undisturbed
main-sequence star. In contrast, the secondary is located far to the
blue of the cluster main-sequence for its luminosity. Also, the observed luminosity
ratio $L_{2}/L_{1}\approx 1/4$ is much too high compared to that
expected for main-sequence stars with $m_{2}/m_{1} < 0.35$.
For stars with $0.43<m<2.0_{\odot}$ we have $L\sim m^{4}$ (Duric 2004). 
If the primary of V8 is indeed an ordinary main-sequence star then 
its mass can be estimated roughly at $0.8~m_{\odot}$ given a cluster 
age of about 13~Gyr. In such a case one obtains $L_{2}/L_{1}=16$ for $q=0.5$, 
a conservative lower limit for V8, as in fact $q<0.35$. 
Clearly it is the the secondary which is the stranger component
of the blue straggler V8.

Such an unusual system may throw some light on problems related to the evolution
and observed status of W UMa-type binaries. The blind acceptance of the contact 
model (Lucy 1968a, 1968b) for very
short period binaries with components of identical effective temperature,
yet very different masses, has been recently strongly contested by the
results for AW UMa (Pribulla \& Rucinski 2008). This flagship case of the
very low mass ratio binary ($q \le 0.1$)  whose light curve so well obeys
the contact model turned out not to be in contact upon a more careful
spectroscopic scrutiny; instead, it appears to be a detached or
semi-detached system engulfed in an optically thick stream with temperature
not different from that of the stellar photosphere. The picture is indeed
confusing, as the control case of V566 Oph ($q = 0.26$) turned out to
agree -- both photometrically and spectrally -- with the contact model.
Apparently, a range in conformance to the contact model exists.

There is no doubt that the present configuration of V8 results from
a substantial mass exchange between the components (it is obvious that 
the secondary component is overluminous given the constraints on
the present mass ratio). Most likely the secondary lost nearly all
matter from its original envelope, and is burning hydrogen in a shell.
Such a situation occurs in some low mass algols (Paczy\'nski 1971; Eggleton 
\& Kisleva-Eggleton 2002). However, before considering any detailed
evolutionary scenario one needs to determine absolute parameters of
the system. This requires supplementing available photometric data with
spectroscopic observations. Obtaining a radial velocity curve for V8 is a
challenging task. The luminosity ratio of $1/4$ is not a big
obstacle, but the low apparent luminosity of the secondary ($V\approx
18.8$) makes the observations  difficult as the short
orbital period of the system limits the exposure time to 15-20
minutes. Useful spectra can be obtained only on the
largest available telescopes.

The spectroscopic masses are essential in relating our current results to
the important finding of Gazeas \& Niarchos (2006) that the masses of secondary
components in W UMa-type binaries are amazingly similar, clustering
around 0.45 $m_\odot$. It is possible that what we normally interpret as
full-size secondaries (filling to different degrees their Roche lobes
to different degrees) are in fact collapsed cores of originally more 
massive components,  as envisaged by Stepien (2006, 2009).

\Acknow{ 
We sincerely thank Slavek Rucinski for his useful comments
on the draft version of this paper. 
Research of JK, MR and KZ is supported by the grant MISTRZ from
the Foundation for the Polish Science and by the grant N N203 379936
from the Ministry of Science and Higher Education. I.B.T.
acknowledges the support of NSF grant AST-0507325. 
Some of the data
presented in this paper were obtained from the Multimission Archive
at the Space Telescope Science Institute (MAST). STScI is operated
by the Association of Universities for Research in Astronomy, Inc.,
under NASA contract NAS5-26555. Support for MAST for non-HST data is
provided by the NASA Office of Space Science via grant NAG5-7584 and
by other grants and contracts. }

%\end{document}

\newpage

% Fig 1
\begin{figure}[t!]
%\centerline{\includegraphics[scale=1.0]{fig1.eps}}
\centerline{\includegraphics[width=120mm]{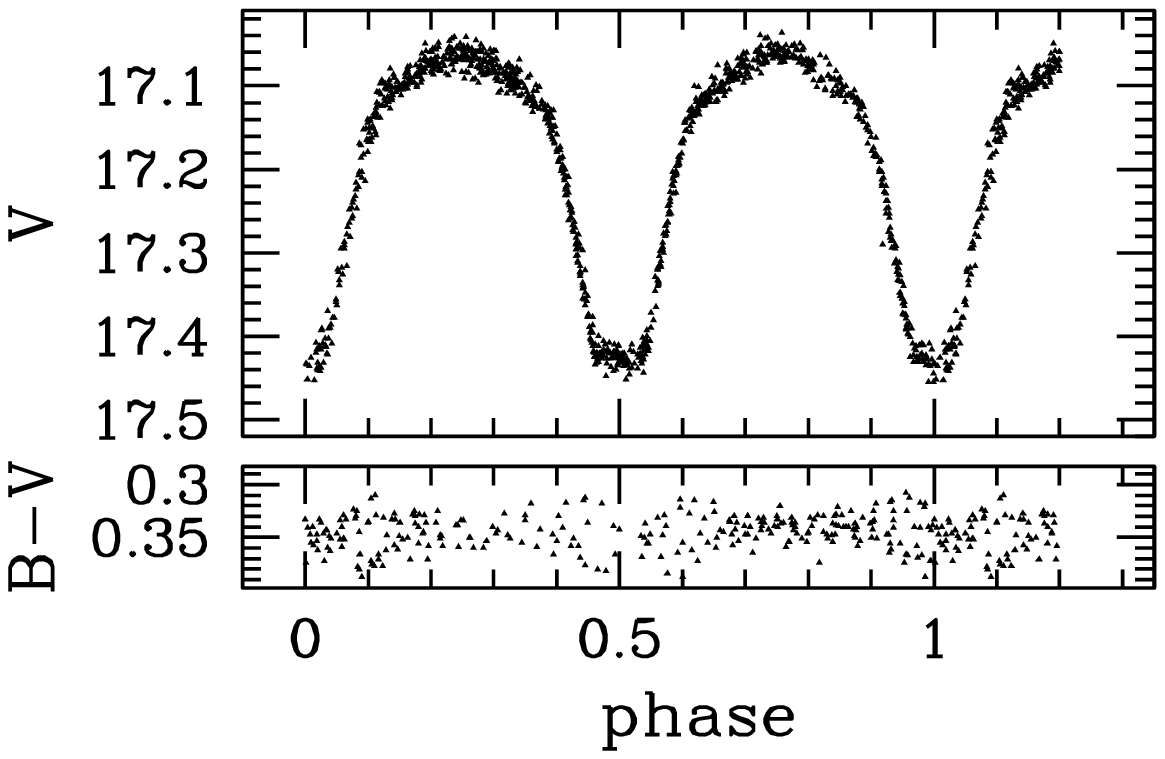}}
\caption{\small Phased light and color curves of V8.}
\end{figure}

% Fig 2
\begin{figure}[t!]
\centerline{\includegraphics[width=120mm]{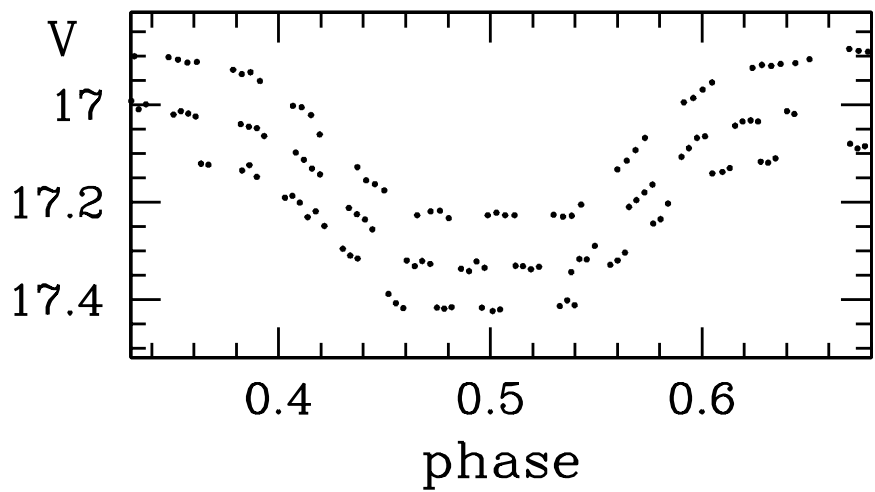}}
\caption{\small Phased light curves of 3 secondary eclipses
of V8 observed in 1998 season. The middle and top curves
were shifted vertically by $-0.1$ and $-0.2$~mag, respectively.}
\end{figure}

% Fig 3
%\begin{figure}[htb]
%\centerline{\includegraphics[width=120mm]{fig3b.ps}}
%\caption{\small
%Left - Reference image  of the region of V8 obtained
%by averaging several images from  1998 season.
%Right - difference of  reference images from 1998 and 2009
%seasons. Positions of two  fast moving stars are marked as well as
%position of the variable V8 (on the right side of field).
%}
%\end{figure}

\begin{figure}[t!]
\includegraphics[width=60 mm, bb = 851 103 1162 414]{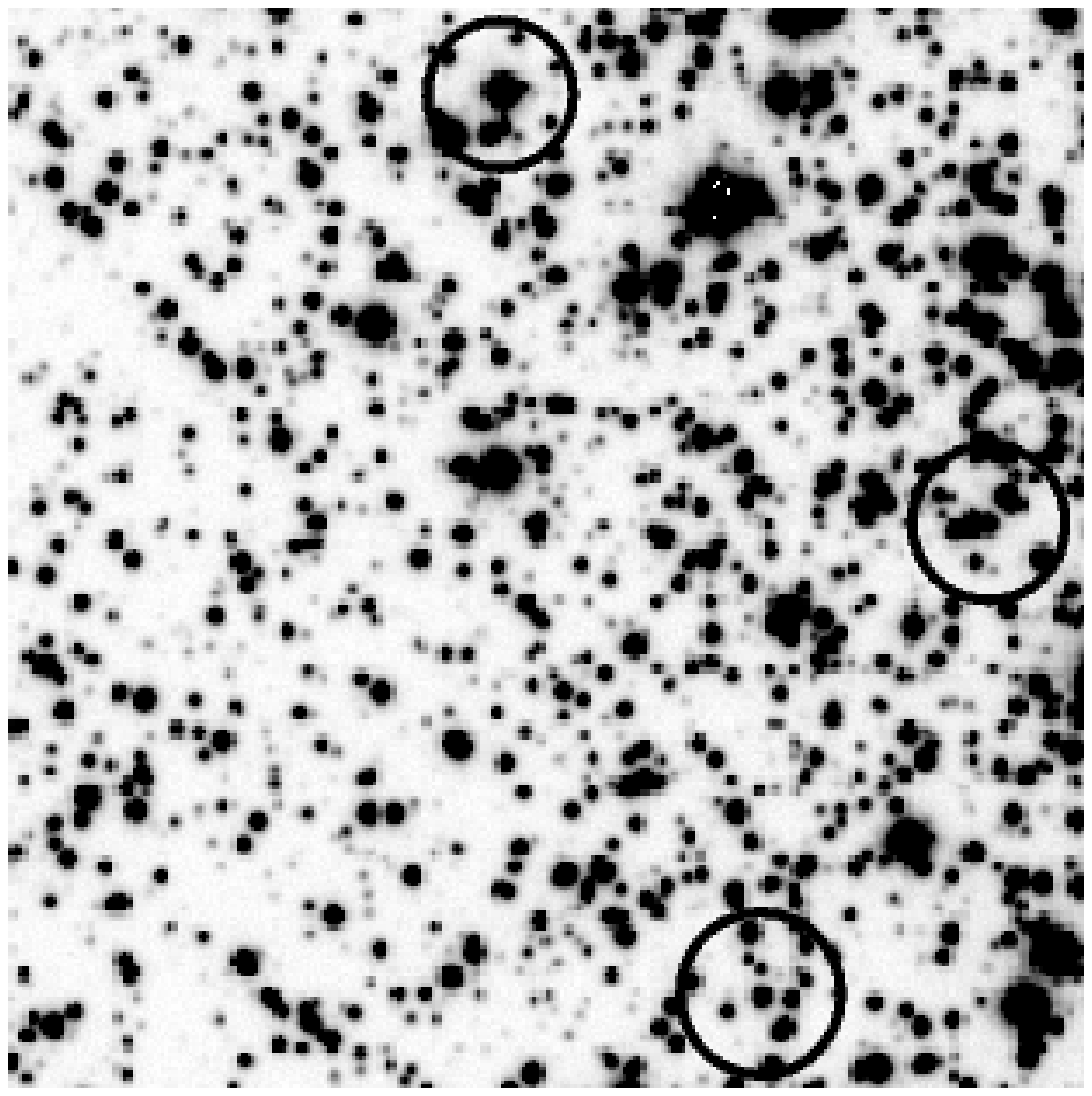}
\includegraphics[width=60 mm, bb = 860 148 1170 458]{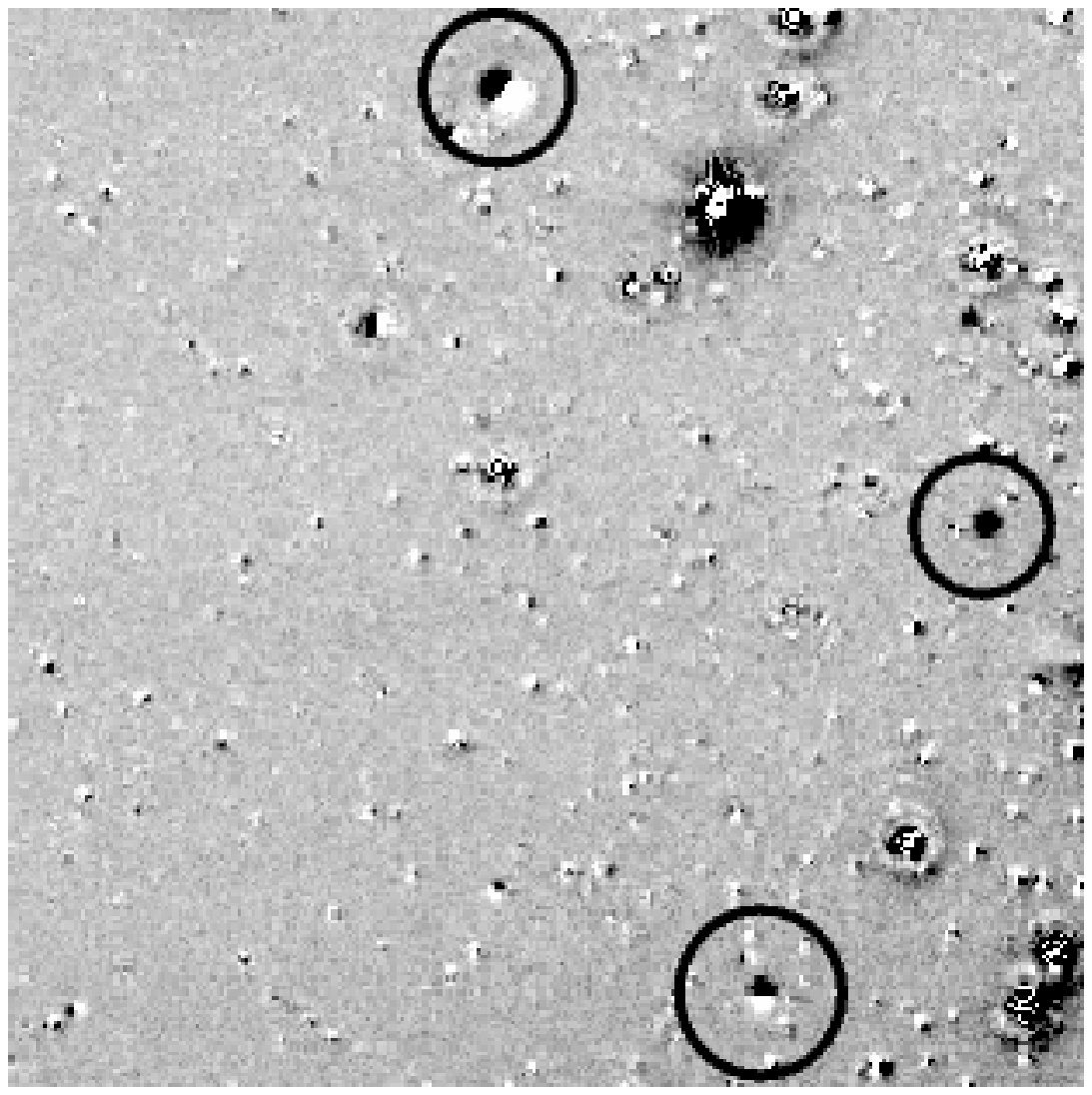}

\caption{\small
Reference image for the neighbourhood of V8, obtained by averaging several 
images taken in 1998 (left) and a residual image obtained by subtracting
the 1998 reference image from the reference image for 2009 data (right). 
Marked are two fast moving stars and V8 itself (at the right 
edge of the field).
}
\label{fig.chart}
\end{figure}

% Fig 4
\begin{figure}[t!]
\centerline{\includegraphics[width=120mm]{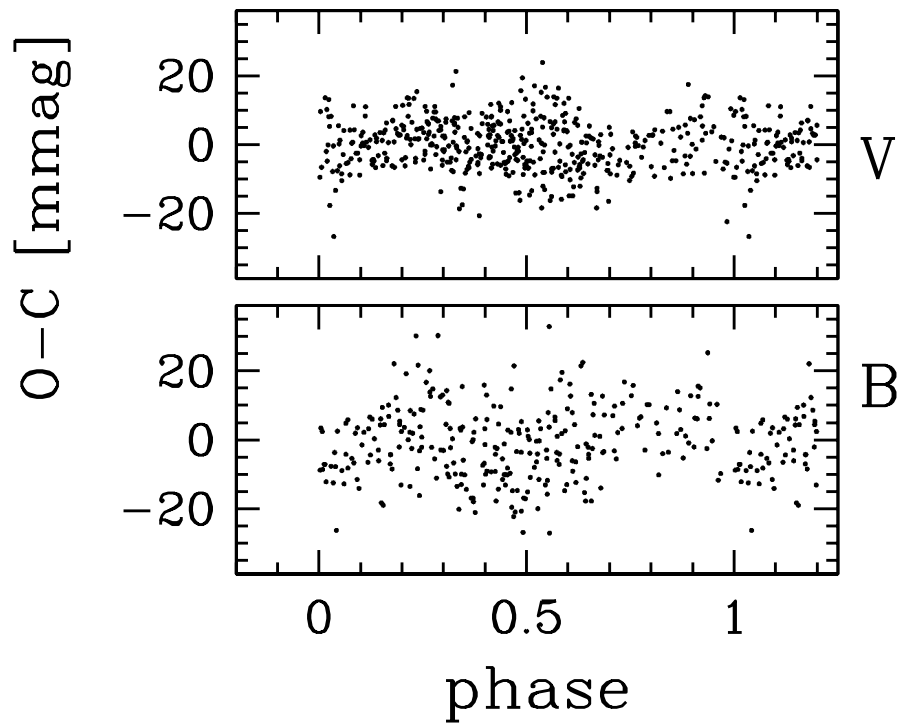}}
\caption{\small
Residuals from the light curve solution for $q=0.20$.
}
\end{figure}

% Fig 5
\begin{figure}[t!]
\centerline{\includegraphics[width=120mm, bb= 59 234 410 605, clip]{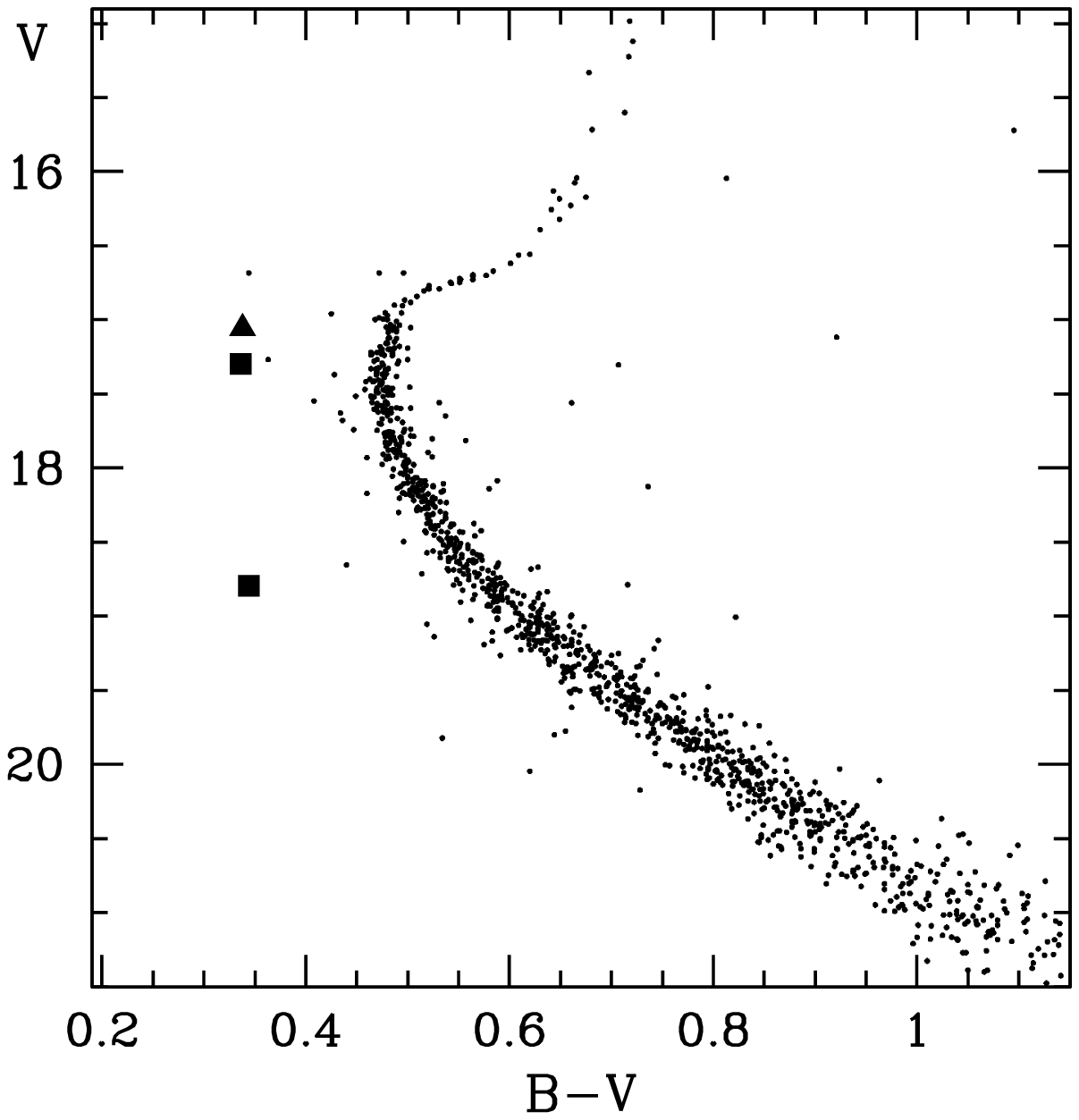}}
\caption{\small
A $V/(B-V)$ color-magnitude diagram of NGC~6752. The components
of V8 are marked by squares, and the triangle shows the 
location of the composite image of the binary. 
}
\end{figure}


\begin{references}

%\refitem{
%Alard, C., \&  Lupton, R.~H.}{1998}{\ApJ}{503}{325}%

%\refitem{
%Alard C.}{2000}{\AA}{144}{363}

\refitem{
Benz, W., \& Hills, J. G.}{1987}{\ApJ}{323}{614}

\refitem{
De Marco, O., Shara, M.~M., Zurek, D., Ouellette, J.~A.,
Lanz, T., Saffer, R.~A., \& Sepinsky, J.~F.}{2005}{\ApJ}{632}{894}

\refitem{
Duric, N.}{2004}{Advanced Astrophysiscs}{Cambridge Univ. Press}

\refitem{	
Eggleton, P.~P., Kiseleva-Eggleton, L.}{2002}{\ApJ}{575}{461}

\refitem{
Eyer, L., \& Wo\'zniak,  P.~ R}{2001}{\MNRAS}{327}{601}

\refitem{
Gazeas, K.~D., Niarchos, P.~G.}{2006}{MNRAS}{370}{L29}

\refitem{
Gilliland, R.~L. et al.}{1998}{\ApJ}{507}{818}


\refitem{
Harris, W.~E.}{1996}{\AJ}{112}{1487}

\refitem{
Hilditch, R.~W., King, D.~J., \& McFarlane, T.~M.}{1988}{\MNRAS}{231}{341}

\refitem{
Hilditch, R.~W., Colier Cameron, A., Hill, G., Bell, S.~A.,
\& Harries, T.~J.}{1997}{\MNRAS}{291}{749}

\refitem{
Kaluzny, J.}{1991}{\Acta}{41}{17}


\refitem{
Kaluzny, J., Rucinski, S.~M., Thompson, I.~B., Pych, W., \&
Krzeminski, W.}{2007a}{\AJ}{133}{2457}


\refitem{
Kaluzny, J., Thompson, I.~B., Rucinski, S.~M., Pych, W.,
 Stachowski, G., Krzeminski, W., \&  Burley, G.~S.}{2007b}{\AJ}{134}{541}


\refitem{
Kaluzny, J., \& Thompson, I.~B.}{2009}{\Acta}{in press}

\refitem{
Knigge, C., Leigh, N., Sills, A.}{2009}{Nature}{457}{288}

\refitem{Kozai, Y.}{1962}{\AJ}{67}{591}

\refitem{Kreiner,  J.~M., Kim,  C.-H., Nha, I.-S.}{2001}{An Atlas
of O-C Diagrams of Eclipsing Binaries, Wydawnictwo Naukowe Akademii
Pedagogicznej, Cracow}{}{}

\refitem{
Kwee, K.~K., \& van Woerden, H.}{1956}{Bull. Astron. Inst. Netherlands}{12}{327}

\refitem{
Lucy, L.~B.}{1968a}{\ApJ}{153}{877}

\refitem{
Lucy, L.~B.}{1968b}{\ApJ}{151}{1123}	


\refitem{
Lucy, L.~B., \&  Wilson, R.~E.}{1979}{\ApJ}{231}{502}

\refitem{
Maceroni, C., Rucinski, S.~M.}{1997}{\PASP}{109}{782}

\refitem{
Mateo,  M.~M., Harris, H.~C., Nemec, J., Olszewski, E.~W.}{1990}{\AJ}
{100}{469}


\refitem{
McCrea, W.~H.}{1964}{\MNRAS}{128}{147}



\refitem{
Mochnacki, S.~W., \& Doughty, N.~A.}{1972}{\MNRAS}{156}{51}


\refitem
{Paczy\'nski B.}{1971}{ARA\&A}{9}{183}


\refitem{
Perets, H.~B., Fabrycky, D.~C.}{2009}{\ApJ}{697}{1048}

\refitem{
Pr\^sa, A., Zwitter, T.}{2005}{\ApJ}{628}{426}

\refitem{
Pribulla, T., Rucinski, S.~M.}{2008}{\MNRAS}{386}{377}

\refitem{
Rucinski, S.~M.}{ 2000}{\AJ}{120}{319}

\refitem{
Rucinski, S.~M.}{1985}{in Interacting binary stars,. eds.
J.~E. Pringle \& R.~A. Wade}{Cambrige Univ. Press}{p. 85}

\refitem{
Rucinski, S.~M., \& Lu, W.}{2000}{\MNRAS}{315}{587}

\refitem{
Sandage, A.~R.}{1953}{\AJ}{58}{61}

\refitem{
Stepien, K.}{2006}{\Acta}{56}{199}

\refitem{
Stepien, K.}{2009}{\MNRAS}{397}{857}


\refitem{
Thompson, I.~B. et al.}{1999}{\AJ}{ 118}{ 462}

\refitem{
Tokovinin A., Thomas S., Sterzik M., Udry, S.}{2006}{\AA}{450}{681}

\refitem
{Wilson, R. E., \& Devinney, E.~J.}{1971}{\ApJ}{166}{605}

\refitem{
Worthey, G.,  \& Lee, H.-c.,}{2005}{preprint}{astro-ph/0604590} 


\end{references}
\end{document}